\documentstyle[11pt]{article}

\def\beq{\begin{eqnarray}}    
\def\eeq{\end{eqnarray}}      

\renewcommand{\Re}{\,\mbox{Re}\,}       

\def\al{\alpha}

\def\ga{\gamma}

\begin{document}


\hfill Preprint numbers: FTUAM/97-2; NF/DF-01/97

\hfill February, 1997

\vskip 5mm

\begin{center}

\setcounter{page}1
\renewcommand{\thefootnote}{\arabic{footnote}}
\setcounter{footnote}0

{\Large \bf On the inflationary solutions in higher-derivative
gravity with dilaton field}

\vskip 8mm
{\bf A.L. Maroto$^a$\footnote{ Electronic address:
maroto@eucmax.sim.ucm.es}
and I.L. Shapiro$^{b,c}$ }
\footnote{ Electronic address: shapiro@fisica.ufjf.br}

\vskip 6mm

{ a) Departamento de F\'{\i}sica Te\'orica,
Universidad Aut\'onoma de Madrid
\\28049, Madrid, Spain}

\vskip 3mm

{ b) Departamento de F\'{\i}sica
-- ICE, Universidade Federal de Juiz de Fora
\\ 33036-330, Juiz de Fora -- MG, Brazil.}
\vskip 3mm

{ c) Department of Mathematical Analysis, Tomsk Pedagogical University
\\ 634041, Tomsk, Russia}

\end{center}

\vskip 8mm

\noindent
\baselineskip 14pt
{\large\sl  Abstract.}
{\it We discuss the existence of de Sitter inflationary solutions
 for the string-inspired fourth-derivative
gravity theories with dilaton field. We consider a space-time of
arbitrary dimension $D$ and an arbitrary parametrization of the 
target space metric.
The specific features of the theory in
dimension $D=4$ and those of the special ghost-free parametrization
of the metric are found. We also consider similar
string-inspired theories
with torsion and construct an inflationary solution with torsion
and dilaton for $D=4$. The stability of the inflationary solutions is also 
investigated.}

\vskip 12mm

\noindent
{\large \bf  Introduction}
\vskip 2mm
\baselineskip 16pt

In the present paper the inflationary cosmological solutions for
the $D$-dimensional metric-dilaton theory of gravity with
action containing second powers of the curvature tensor
are considered. The interest of including into the
action  terms with higher derivatives \cite{ste} 
is due to the fact that they
naturally arise in the string effective action (see, for example,
\cite{gsw}), and also when quantum corrections to the Einstein
action are generated by the conformal anomaly of quantized matter fields
on curved background \cite{anom}. In both cases
the effective action contains, along with the metric, an additional
scalar field called dilaton. The general fourth derivative metric-dilaton
action which interpolates between string-inspired and anomaly-inspired
particular actions is too cumbersome \cite{aeli}, and that is why
here we restrict our study to the string-inspired higher-derivative
models\footnote{One inflationary solution for the anomaly-induced
action for the theory with torsion has been obtained in the third
reference in \cite{anom} (see also \cite{book})}.
Such an action doesn't have higher derivatives in the
dilaton sector (throughout the paper we
use Minkowski signature, notation
$R^{\lambda}_{\;\;\mu\nu\rho}=\partial_{\rho}
\Gamma^{\lambda}_{\mu\nu}-\dots$ and parametrization of \cite{mets}).
Let us first consider the torsionless case, then the mentioned effective
action can be written as
\beq
S_M=\frac{2}{\kappa^2}\int d^Dx \sqrt{g}\; e^{-2\phi}\,
\left\{ -R + 4\,(\partial
\phi)^2+ \alpha'\,\left( a_1R_{\lambda\mu\nu\rho}R^{\lambda\mu\nu\rho}
+
a_2R_{\mu\nu}R^{\mu\nu}+a_3R^2\right)\right\}
\label{1e}
\eeq
The values of the dimensionless parameters $a_1,\,a_2,\,a_3$
have been calculated
for bosonic and heterotic string \cite{ket}
whereas for the superstring the fourth-derivative terms are absent.
In fact the coefficients $a_2,\,a_3$ derived from
string theory contain an arbitrariness related with the possibility
of performing the reparametrization of the background
metric $g_{\mu\nu}$
\beq
g_{\mu\nu} \longrightarrow
g'_{\mu\nu} =
g_{\mu\nu} +
\al '\left(x_1\,R_{\mu\nu} + x_2\,R\, g_{\mu\nu}\right) + ...
\label{repar}
\eeq
where $x_{1,2}$ are  arbitrary parameters.
In particular in any space-time (target-space) dimension,
one can choose some special
parametrization in which only the massless graviton field is
propagating in the spin-two sector, whereas the massive unphysical
ghosts are absent
\cite{zwei,dere}. For such a parametrization
the curvature squared terms appear in the combination:
$$
E = R_{\lambda\mu\nu\rho}\,R^{\lambda\mu\nu\rho} -
4\,R_{\mu\nu}\,R^{\mu\nu}+ R^2
$$
which, in $D=4$, is the integrand of the Gauss-Bonnet topological
term. For $D\neq 4$ this term is not topological but it still
doesn't contribute to the spin-2 massive pole in the propagator.
The propagation of the spin-0 massive states, however, depends on the
decomposition of the metric $g_{\mu\nu}$.
If one separates the conformal factor of the metric, then the massive
spin-0 pole appears. Such a pole, indeed, depends on the gauge fixing and
doesn't violate unitarity.

On the other hand, from the
string theory point of view such a special parametrization can not be
distinguished from the others \cite{tse,ovrut},
and that is why it is interesting to explore
the nonsingular inflationary solutions for
the general action (\ref{1e}). As it was recently demonstrated
in \cite{highderi}, the renormalization
properties of the effective theory of gravity theory are very different
in the special ghost-free parametrization of the metric and in the general
one, in which all the ghosts propagate. The two theories are, however,
equivalent in the low-energy IR limit \cite{don,highderi}.
Inflation is another interesting aspect
in which the properties of the effective string-inspired gravity
may depend (or be independent) on the parametrization. The existence
of the inflationary solutions can be introduced
in order to constraint the reparametrization arbitrariness
in the string effective action, since it would be desirable to have a
period of inflation as a natural consequence of string theory at
energies below (but not far below) Planck scale. At these
energies the first higher derivative corrections to the low-energy
string effective action can be relevant.

Recently there has been an extensive study of the
(inflationary) cosmological solutions
for the string-inspired gravitational theory with higher derivatives
restricted by the mentioned ghost-free parametrization of the metric
(see \cite{cosm1,bebe} and references therein)
with $- (1/4)\,a_2 = a_3 = a_1$. The special case
of $a_1=a_2=0$ was also explored \cite{barrow}. In the context of the
effective field theory for gravity, the cosmological
solutions in the dilatonless case obtained from two-loop pure
quantum gravity and integrating out conformal free matter
in the Standard Model have also been studied \cite{cosm2}.
The existence of the de Sitter solutions and the boundary 
conditions in the presence of the
generic higher order terms (but without dilaton or torsion)
was discussed in \cite{barrowmadsen} (see also references therein).
In the present article we consider the possibility of the inflationary
conformally flat solutions in the general theory (\ref{1e}),
and therefore start from
arbitrary nonrestricted values of $a_1,\,a_2,\,a_3$, we also explore the 
same problem for the theory with torsion.

The paper is organized in the following way. In section 2 the
equations of motion for the general theory of the type (\ref{1e})
are derived. In sections 3 and 4 we study the
inflationary solutions for theories without and with dilaton field.
In section 5 the generalization of (\ref{1e}) for the theory with torsion
is considered. Section 6 is devoted to the study of the stability of the
de Sitter solutions. The last section contains some conclusions.

\vskip 6mm
\noindent
{\large \bf 2.$\,$ D-dimensional space-time and general
equations of motion}
\vskip 2mm
\baselineskip 16pt

Consider D-dimensional
 $k=0$ (conformally flat) Friedmann-Robertson-Walker (FRW) metric:
\beq
ds^2=-dt^2+a(t)^2\,d^{(D-1)}\,X^2
\label{2}
\eeq
where $d^{(D-1)}\,X^2$ denotes the $(D-1)$-dimensional flat space metric.
It is useful to define $\,b(t)=\log a(t)\,$
which implies $\,\dot b=\dot a/a=H(t)$, with $H$ being the
Hubble parameter. Taking into account
that in an isotropic and homogeneous space-time the dilaton
field can only depend on $t$, we rewrite the action integral (\ref{1e})
as follows:
$$
S_M \,= \,  \int dt \,e^{(D-1)b}\,e^{-2\phi}\,\left\{
(D-1)(H^2 D+2\dot H) - 4 \,\dot \phi^2 +
\right.
$$$$
\left.
+ \;\alpha'(D-1)\,\left[2\,a_1\left( 2\,\dot H^2+
4\,H^2\,\dot H+D\,H^4 \right)
\right. \right.
$$
\beq
\left.\left.
+ a_2\,\left(D\dot H^2+(D-1)DH^4+4(D-1)\dot H H^2\right)
+ a_3\,(D-1)\,(H^2\,D+2\dot H)^2  \right] \right\}
\label{le}
\eeq
Since we will deal with different versions of the last
action, it is convenient
to write the equations of motion for the scale factor and the
dilaton field for the generic action of the
form:
\beq
S=\int \,dt \,e^{(D-1)b}\,e^{-2\phi}\,L(H,\dot H,\dot \phi)
\label{3}
\eeq
They are:
\begin{eqnarray}
&(D-1)&L-(D-1)H\frac{\partial L}{\partial H}
+2\dot \phi\frac{\partial L}{\partial H}
-\frac{d}{dt}\frac{\partial L}{\partial H}
+(D-1)^2H^2\frac{\partial L}{\partial \dot H}\nonumber \\
-4&(D-1)&H\dot \phi\frac{\partial L}{\partial \dot H}
+2(D-1)H\frac{d}{dt}\frac{\partial L}{\partial \dot H}
+4\dot\phi^2\frac{\partial L}{\partial \dot H}
-2\dot\phi\frac{d}{dt}\frac{\partial L}{\partial \dot H}\nonumber \\
+&(D-1)&\dot H \frac{\partial L}{\partial \dot H}
-2\ddot \phi\frac{\partial L}{\partial \dot H}
-2\dot \phi\frac{d}{dt}\frac{\partial L}{\partial \dot H}
+\frac{d^2}{dt^2}\frac{\partial L}{\partial \dot H}=0
\label{4}
\end{eqnarray}
and
\begin{eqnarray}
-2\,L-(D-1)\,H\,\frac{\partial L}{\partial \dot\phi}
+2\,\dot\phi\,\frac{\partial L}{\partial \dot\phi}
-\frac{d}{dt}\,\frac{\partial L}{\partial \dot\phi}=0
\label{5}
\end{eqnarray}

In the next sections the equations (\ref{4}) and (\ref{5}) will
be applied to the theory (\ref{le}) with constant and dynamical
dilaton field.
\vskip 6mm
\noindent
{\large \bf 3.$\,$ Solution without dilaton}

\vskip 2mm
\baselineskip 16pt

 We now return to the low-energy string effective action (\ref{le})
and consider the simple case with constant dilaton field.
Then the action integral can be written as follows:
\begin{eqnarray}
S_M = \int dt\, e^{(D-1)b}\,\left\{
(D-1)\,(H^2\,D+2\,\dot H)+\alpha'\,
\left[
\gamma_1\,\dot H^2+\gamma_2\,H^2\,\dot H+\gamma_3\,H^4 \right]\,\right\}
\label{6}
\end{eqnarray}                                     
where:
\begin{eqnarray}
\gamma_1&=&4\,a_1\,(D-1)+a_2\,D\,(D-1)+4\,a_3\,(D-1)^2 \nonumber \\
\gamma_2&=&8\,a_1\,(D-1)+4\,a_2\,(D-1)^2+4\,a_3\,D\,(D-1)^2 \nonumber \\
\gamma_3&=&\frac{D}{4}\,\gamma_2
\label{gammas}
\end{eqnarray}

The equation of motion for the scale factor reads in this case:
\begin{eqnarray}
&\,&H^2(D-1)^2(D-2)+2\dot H(D-1)(D-2)
+\alpha'\left(3{\dot H}^2(D-1)\gamma_1 \right.\nonumber \\
&+& \left. H^2\dot H(2(D-1)^2\gamma_1+4(D-1)\gamma_2-12\gamma_3)
+H^4(-3(D-1)\gamma_3+(D-1)^2\gamma_2)\right. \nonumber \\
&+& \left. 4H\ddot H (D-1)\gamma_1
+2\frac{d}{dt}\ddot H\gamma_1\right)=0
\label{7}
\end{eqnarray}

We look for (anti-)de Sitter solutions of the above equation.
In terms of the Hubble
parameter an (anti-)de Sitter solution is simply $H=H_0=
constant$
(with our notation de Sitter space-times
have constant negative curvature, i.e.
$R=-(D-1)DH_0^2<0$ which implies $H_0^2>0$, anti-de Sitter
space-times have positive curvature),
then the above equation reduces to the
algebraic one:
\beq
H_0^2\,(D-1)^2(D-2)
+\alpha'\,H_0^4\,{\gamma_3}\,\frac{(D-4)(D-1)}{D} = 0
\label{8}
\eeq

The first important point is that this equation does not depend on
$\gamma_1$ but only on $\gamma_{2,3}$. Such
independence has direct sense.
One can rewrite the curvature squared terms of the starting action
in another basis
$$
a_1\,R_{\lambda\mu\nu\rho}^2 + a_2\,R_{\mu\nu}^2 + a_3R^2 =
$$
\beq
= - \,\frac{4\,a_1 + (D-2) \, a_2}{4\,(D-3)}\;E +
\frac{D-2}{D-3}\;\left( a_1 + \frac{a_2}{4}\right)\,C^2 +
\frac{\ga_1}{4\,(D-1)^2}\;R^2
\label{in1}
\eeq
where
$$
C^2 = R_{\lambda\mu\nu\rho}^2 - \frac{4}{D-2}\,R_{\mu\nu}^2 +
\frac{2}{(D-1)(D-2)}R^2
$$
is the square of the Weyl tensor. Indeed for the above FRW metric
$C^2 = 0$ and hence only two combinations of the coefficients $a_{1,2,3}$
should be relevant. On the other hand, the $\gamma_1$ -- independence
of the equations
doesn't indicate that for the given metric the equations are
completely independent on the presence of the $R^2$ term,
because $\ga_{2,3}$ is not proportional to the coefficient of the
Gauss-Bonnet term $E$ in (\ref{in1}). This indicates that the existence
of the inflationary de Sitter solutions can, in principle, depend on the
choice of parametrization (\ref{repar}) of the "target-space" metric
$g_{\mu\nu}$ and also on the space-time dimension.  In four
dimension not only the $C^2$ term is absent, but the Gauss-Bonnet
term $E$ contributes as a total derivative to the action (in the absence
of the dilaton field) and accordingly only the term proportional to
$\gamma_1R^2$ would contribute to the equations of motion. Since the
(anti-)de Sitter solutions are independent on $\gamma_1$ this
explains that in
$D=4$ there is no (anti-)de Sitter solutions different from
the trivial Minkowski one $H_0=0$. For arbitrary dimension
the solutions
only depend
on the combination $\gamma_3=2a_1D(D-1)+a_2(D-1)^2D+a_3(D-1)^2D^2\,$
and they can be written as:
\begin{eqnarray}
H_0&=&0 \nonumber \\
H_0^2&=&-\,\frac{D(D-1)(D-2)}{\alpha'\,\gamma_3\,(D-4)}
\label{sds}
\end{eqnarray}

Notice again that four dimensions is the only (apart from $D=2$) case in
which
there is no (anti-)de Sitter solution independently of the
parametrization. The sign of $H_0^2$ (and therefore the type of solution)
depends on the sign of $\gamma_3$, and thus can be affected by
reparametrizations of the target-space metric (\ref{repar}).
\vskip 6mm
\noindent
{\large \bf 4.$\,$ Solutions with dilaton}

\vskip 2mm
\baselineskip 16pt

In the theory with dynamical dilaton the low-energy string
effective action (\ref{le}) can be rewritten in a form similar to (\ref{6})
\beq
S_M = \int dt\, e^{(D-1)b}\,e^{- 2\phi}\,\left\{(D-1)(H^2D+2\dot H)
-4\dot\phi^2 + \alpha'
\left(\gamma_1\dot H^2+\gamma_2H^2\dot H+
\gamma_3H^4\right) \right\}
\label{9}
\eeq
The corresponding equations of motion for the scale factor
and dilaton are:
\begin{eqnarray}
&\,&H^2(D-1)^2(D-2)+2\dot H(D-1)(D-2)
+\alpha'\left(3{\dot H}^2(D-1)\gamma_1 \right.\nonumber \\
&+& \left. H^2\dot H(2(D-1)^2\gamma_1+4(D-1)\gamma_2-12\gamma_3)
+H^4(-3(D-1)\gamma_3+(D-1)^2\gamma_2)\right. \nonumber \\
&+& \left. 4H\ddot H (D-1)\gamma_1
+2\frac{d}{dt}\ddot H\gamma_1\right)
-4\dot \phi H(D-1)(D-2)+4{\dot \phi}^2(D-1)-4\ddot \phi(D-1) \nonumber \\
&+&\alpha'\left(\dot \phi H \dot H (-4\gamma_2-8(D-1)\gamma_1)
-8\frac{(D-2)}{D}\gamma_3\dot \phi H^3
+8{\dot \phi}^2\dot H\gamma_1 \right.\nonumber \\
&+&\left. 4\gamma_2{\dot \phi}^2H^2
-8\dot \phi \ddot H\gamma_1-4\gamma_1\ddot\phi\dot H
-2\gamma_2\ddot \phi H^2\right)=0
\label{10}
\end{eqnarray}
and
\beq
(D-1)(H^2 D + 2 \dot H - 4 H \dot\phi)+
4\dot \phi^2 - 4\ddot \phi
+ \alpha'(\gamma_1\dot H^2 + \gamma_2 H^2\dot H + \gamma_3 H^4) = 0
\label{11}
\eeq
It is easy to see that the above equations may have
flat Minkowski space-time as a solution. In this case $H=0$
and then both equations above reduce to the same equation for the dilaton:
\beq
-\dot\phi^2+\ddot \phi=0
\label{12}
\eeq
whose solutions are:
\begin{eqnarray}
\phi &=&\mbox{constant}\nonumber \\
\phi &=&-\,\mbox{log}\left(\frac{t-t_0}{\tau}\right)
\label{13}
\end{eqnarray}
where $t_0$ and $\tau$ are integration constants. The $\phi=$constant
case agrees
with the Minkowski solution found in the previous section.

Equations (\ref{10}), (\ref{11}) are very cumbersome and
it is rather difficult to get  solutions different from the flat one.
As in the previous case, let us try to find the solutions
 with $H=H_0={constant}$, that is (anti-)de Sitter space-times.
For simplicity we will also assume a linear dilaton, i.e,
$\dot\phi=v={constant}$.
In this case
the equations of motion (\ref{10}) and (\ref{11})
become algebraic equations for two unknowns:
\begin{eqnarray}
(D-1)^2(D-2)H_0^2 &+& \alpha' \,\gamma_3\,\frac{(D-4)(D-1)}{D}\,H_0^4
-4\,(D-1)(D-2)v\,H_0 \nonumber \\
- 8\alpha'\,v\,\frac{D-2}{D}\,\gamma_3\,H_0^3
&+& \frac{16}{D}\,\alpha'\, v^2\,\gamma_3\,H_0^2 + 4\,(D-1)\,v^2=0
\label{14}
\end{eqnarray}
 and
\begin{eqnarray}
-4\,v^2+4\,v\,H_0\,(D-1)-\,(D-1)\,D\,H_0^2 - \,\alpha'\,\gamma_3\,
H_0^4=0
\label{15}
\end{eqnarray}

{}From these equations we see that the solution can only depend on
$\gamma_3$, just
as in the case without dilaton. The general theorems of algebra ensure
that the above higher order system of algebraic equations
has some solutions, but it is not explicit that
they are real. The analysis of the system shows that there can be
three real exact solutions:

i) the trivial one $H=0$, $v=0$ and

ii) the nontrivial ones
\begin{eqnarray}
H_0 &=&\pm\,\sqrt{\frac{1-D}{\alpha'\,\gamma_3}}
\nonumber \\
v &=& \frac{(D-1)}{2}\,H_0
\label{16}
\end{eqnarray}
As in the previous case the sign of $H_0^2$ depends on the sign of
$\gamma_3$.
The above pair of non-trivial solutions are related by the lowest-order
scale factor duality \cite{Vene} given by:
\begin{eqnarray}
H &\rightarrow & -H \nonumber \\
v &\rightarrow & v-(D-1)H
\label{dual}
\end{eqnarray}
 Notice also that the presence of a dynamical dilaton modifies the
solutions in (\ref{sds}).
In addition to this simple solution there are three pairs of
complex conjugated solutions.
The expression for the complex solutions cannot be given in a
compact form for
arbitrary dimension, in addition they
yield  complex values for the dilaton,
thus we disregard to present their explicit form here.

\vskip 6mm
\noindent
{\large \bf 5.$\,$ Solutions with torsion}

\vskip 2mm
\baselineskip 16pt

The low-energy string effective action contains in addition to
the graviton and dilaton fields an antisymmetric tensor field
$H_{\mu\nu\lambda}$, usually referred to as the string torsion.
The explicit expression for the effective action in this case
is given by \cite{mets}:
\beq
S_M &=&\frac{2}{\kappa^2}\int d^Dx \,\sqrt{g} e^{-2\phi}
\left\{ -R +4(\partial \phi)^2+\kappa_0 H_{\alpha\beta\gamma}^2 +
\alpha'\,\left( a_1R_{\lambda\mu\nu\rho}
R^{\lambda\mu\nu\rho}\right.\right.\nonumber \\
&+&
a_2R_{\mu\nu}R^{\mu\nu}+a_3R^2
+ \kappa_1R^{\alpha\beta\rho\sigma}
H_{\alpha\beta\lambda}H_{\rho\sigma}^{\;\;\;\lambda}\nonumber \\
&+&\left.\left.
\kappa_2H_{\mu\nu\lambda}H^\nu_{\;\;\rho\alpha}H^{\rho\sigma\lambda}
H_{\sigma}^{\;\;\mu\alpha}
+\kappa_3H_{\mu\alpha\beta}H_\nu^{\;\;\alpha\beta}H^{\mu\rho\sigma}
H^{\nu}_{\;\;\rho\sigma}\right)\right\}
\label{1}
\eeq

 Taking into account the antisymmetry of $H_{\alpha\beta\gamma}$,  we
can write $\,H_{\alpha\beta\gamma}=
\epsilon_{\alpha\beta\gamma\rho}\,S^{\rho}\,$. For a FRW
space-time and assuming also the
homogeneity and isotropy for the $S_{\mu}$
pseudo-vector one has to choose \cite{book} $\,S_{\mu}=(T(t),0,0,0)\,$,
then the action (\ref{1}) reads:
$$
S_M =  \int dt \,e^{(D-1)b} e^{-2\phi}\,\left\{
(D-1)(H^2 D+2\dot H) -4\dot\phi^2 +b_0 T^2
\right.
$$$$
\left.
+ \alpha'\;(D-1)\,\left[2\,a_1\left( 2\,\dot H^2+
4\,H^2\,\dot H+D\,H^4 \right)  +
\right. \right.
$$$$
\left.\left.
+ a_2\,\left(D\dot H^2+(D-1)DH^4+4(D-1)\dot H H^2\right)
+ a_3\,(D-1)\,(H^2\,D+2\dot H)^2 \right. \right.
$$
\beq
\left.\left.
+b_1T^2H^2+b_2T^4+b_3T^4 \right] \right\}
\label{let}
\eeq
where
$$b_0=\kappa_0(D-1)(D-2)(D-3)$$
$$b_1=-2\kappa_1(D-3)(D-2)(D-1)$$
$$ b_2=\kappa_2(D-1)(D-2)(D-3)^2$$
$$b_3=\kappa_3(D-1)(D-2)^2(D-3)^2$$

The equation of motion for torsion reads:
\begin{equation}
2\,b_0\,T+2\,b_1\,\alpha'\,T\,H^2+4\,\alpha'\,(b_2
+b_3)\,T^3 = 0
\label{torsi}
\end{equation}
and
since the action does not contain derivatives of the torsion, it is an
algebraic equation.
The corresponding solutions are: $T=0$ and
\beq
T^2= -\,\frac{b_0+b_1\,\alpha'\,H^2}{2\,\alpha'\,(b_2+b_3)}
\label{torsion}
\eeq
the first one
when substituted back into the action leads to the solutions for metric
and dilaton presented
in the previous section. The second solution (\ref{torsion})
for torsion when
substituted gives rise to:
\beq
S_M=\int dt \,e^{(D-1)b}e^{-2\phi}\,
\left\{ c_0 -4\dot\phi^2+c_1H^2+c_2\dot H+c_3H^2\dot H+c_4\dot H^2+
c_5H^4\right\}
\label{24}
\eeq
where
$$
c_0=-\frac{b_0^2}{4\alpha'(b_2+b_3)}
,\,\,\,\,\,\,\,\,\,\,\,\,\,
c_1=(D-1)D-\frac{b_0b_1}{2(b_2+b_3)}
,\,\,\,\,\,\,\,\,\,\,\,\,\,
c_2=2(D-1)
$$$$
c_3=\alpha'\gamma_2
,\,\,\,\,\,\,\,\,\,\,\,\,\,
c_4=\alpha'\gamma_1
,\,\,\,\,\,\,\,\,\,\,\,\,\,
c_5=\alpha'\left[\gamma_3-\frac{b_1^2}{4(b_2+b_3)}\right]
$$
Therefore the effect of a non-vanishing torsion is to modify the
coefficients in
(\ref{le}) and introduce a cosmological constant
term $c_0$. Let us first consider
the simple
case in which the dilaton is absent. The
equation of
motion for the scale factor can be obtained directly from (\ref{4})
and the corresponding (anti-)de Sitter solutions read:
\beq
H_0^2=\frac{A \pm \sqrt{A^2-2(D-1)(D-2)B\kappa_0^2}}{\alpha'B}
\label{solution}
\eeq
where: $A=-4(D-1)(D-2)(\kappa_0\kappa_1-\kappa_2-(D-2)\kappa_3)$ and
 $B=-24\kappa_1^2(D-1)(D-2)-8\kappa_2\gamma_3(D-4)/D
-8\kappa_3\gamma_3(D-2)(D-4)/D$.
A particular case corresponds to the $\sigma$-parametrization in which
for the bosonic string
$$
\kappa_0=-1/12,
\,\,\,\,\,\,\,\,\,\,\,\,\,
\kappa_1=1/8,
\,\,\,\,\,\,\,\,\,\,\,\,\,
\kappa_2=1/96,
\,\,\,\,\,\,\,\,\,\,\,\,\,
\kappa_3=-1/32
$$
while
$$
\gamma_1=(D-1),
\,\,\,\,\,\,\,\,\,\,\,\,\,
\gamma_2=2(D-1),
\,\,\,\,\,\,\,\,\,\,\,\,\,
\gamma_3=D(D-1)/2
$$
For these parameters we obtain
the following solutions valid for different dimensions.
\vskip 1mm

For $D=2,3$ there is no solution since in that case $T\neq 0$ is not a
solution of  (\ref{torsi}).
\vskip 1mm

For $D=4$,  $\,\,\alpha'H_0^2=(4\pm \sqrt{19})/9$. Since one of
the above solutions is
positive, in this dimension we have inflationary de Sitter solution
which appears due to the torsion terms.
\vskip 1mm

For $D=5$, $\,\,\alpha'H_0^2=(42\pm 4\sqrt{115})/38$,
one of the solutions is positive.
\vskip 1mm

For $D=10$, $\,\,\alpha'H_0^2=(-88\pm 10\sqrt{77})/33$
both solutions are negative.
\vskip 1mm

For $D=26$, $\,\,\alpha'H_0^2=(-840\pm 2\sqrt{175727})/673$
both solutions are negative.
\vskip 1mm

Thus the existence of the acceptable inflationary solutions for the
theory
with torsion strongly depends on the dimension of the space-time. In
particular
such a solution exists for $D=4$ in this parametrization.

In the general case with dynamical dilaton and torsion fields and for an
arbitrary parametrization,
the equations of motion can again be
obtained from (\ref{4}).
The explicit form of the equations is very cumbersome and that is why
we present only final results.
The corresponding maximally symmetric solutions
are the following:

i) Two flat Minkowski solutions with a linear dilaton given by:
\begin{eqnarray}
H_0 &=&0 \nonumber \\
v &=&\pm\frac{\kappa_0 \sqrt{(D-1)(D-2)}}{4\sqrt{\alpha'\Lambda}}
\end{eqnarray}
where $\Lambda=\kappa_2+(D-2)\kappa_3$. We remark that for the given
$k_{2,3}$ this solution (for $v$)
becomes complex for $D>2$, so we present it here for generality only.

ii) Two pairs of (anti-)de Sitter solutions:
\begin{eqnarray}
H_0^2 &=&
\frac{\kappa_0\kappa_1(D-2)(D-1)+(D-1)\Lambda}
{2\alpha'\left(\kappa_1^2(D-2)(D-1)-\gamma_3\Lambda\right)}
 \nonumber \\
&\pm&\frac{\sqrt{(D-1)\Lambda \left(\gamma_3\kappa_0^2(D-2)
+2\kappa_0\kappa_1(D-2)(D-1)+(D-1)\Lambda\right)}}{
2\alpha'\left(\kappa_1^2(D-2)(D-1)-\gamma_3\Lambda\right)}\nonumber \\
v &=&\frac{(D-1)}{2}H_0
\label{solu2}
\end{eqnarray}

As in the torsionless case each pair (with opposite signs in $H_0$) of
solutions are related by the
duality transformations given in (\ref{dual}). Notice also that
these solutions only depend
on the $\gamma_3$ coefficient (\ref{gammas}) of the curvature terms.
In the case of the $\sigma-$parametrization for
the bosonic string, it can be shown that independently on
the dimension only anti-de Sitter solutions arise.

\vskip 8mm
\noindent
{\large \bf 6.$\,$ Stability of the solutions}
\vskip 2mm
\baselineskip 16pt

In order to study the stability of the above obtained de Sitter solution we 
will consider small perturbations around them given by:
\begin{eqnarray}
H(t)=H_0+\delta(t),\;  \;  \;  \;  \;  \;  \;  
\dot \phi(t)=v+\epsilon(t)
\end{eqnarray}
One has to notice that while the 
torsion is taken in the form (\ref{let}), its perturbations 
are not relevant. After 
solving the equations (\ref{torsi}) such  perturbations reduce 
to  modifications in $\delta(t)$ and $\epsilon(t)$.

Following the general calculational method of section 2,
we consider the general theory with action 
(\ref{3}) and introduce the notations:
\begin{eqnarray}
\begin{array}{ccc}
L_1=\left.\frac{\partial L}{\partial H}\right\vert_0 & 
L_2=\left.\frac{\partial L}{\partial \dot H}\right\vert_0 &
L_3=\left.\frac{\partial L}{\partial \dot \phi}\right\vert_0 \\
L_{11}=\left.\frac{\partial^2 L}{\partial H^2}\right\vert_0 &
L_{12}=L_{21}=\left.\frac{\partial^2 L}{\partial H\partial \dot H}
\right\vert_0 &  
L_{13}=L_{31}=\left.\frac{\partial^2 L}{\partial H\partial \dot \phi}
\right\vert_0\\
L_{22}=\left.\frac{\partial^2 L}{\partial \dot H^2}\right\vert_0 &
L_{23}=L_{32}=\left.\frac{\partial^2 L}{\partial \dot H\partial\dot \phi}
\right\vert_0 &
L_{33}=\left.\frac{\partial^2 L}{\partial \dot \phi^2}\right\vert_0 
\end{array}
\end{eqnarray}
where the $\,0\,$ subindex indicates that the derivatives are taken at the 
point of extremal  $\,\,H=H_0,\, \;\dot H=0, \;\, \dot\phi=v\,\,$.
Below we only consider  real values for $H_0$ and $v$.

Linearizing the equations (\ref{4}) and (\ref{5}) in 
the perturbations $\delta$ and $\epsilon$ we obtain
the following two  equations: 
$$
\delta \left[
- L_{11}K+L_{12}K^2+2(D-1)L_2 K\right]
+\dot \delta\left[
2(D-1)L_2+L_{12}K+L_{22}K^2-L_{11}\right]
+2 \ddot \delta L_{22}K +
$$
\beq
+\frac{d}{dt}\ddot \delta L_{22}+
\epsilon\left[(D-1)L_3-L_{13}K+L_{23}K^2-4L_2 K\right]
+\dot \epsilon \left[
-L_{13}-2L_2-2L_{23}K\right] +\ddot \epsilon L_{23} = 0
\label{stab}
\eeq
\beq
 \delta\left[ 2L_1+(D-1)L_3+L_{13}K\right]
+\dot\delta \left[ 2L_2 +L_{13}+L_{23}K\right]
+\ddot \delta L_{23}+\epsilon L_{33}K +\dot \epsilon L_{33}=0
\label{stab1}
\eeq
where $K=(D-1)H_0 -2v$. In the simple case without dilaton field, only the 
first of these equations is needed. This is a third order algebraic 
equation, and one can analise its solution by standard methods. 
In order to simplify this we
notice that the equation can be rewritten in the more simple form 
\begin{eqnarray}
\dot {\cal F}+K{\cal F}=0
\label{char}
\end{eqnarray}
where $\,{\cal F}=\ddot \delta L_{22}+K\dot \delta L_{22}+\delta(L_{12}K
+ 2 L_2 (D-1) - L_{11})\,$
and $\,K=(D-1)H_0\,$. The characteristic equation of 
(\ref{stab}) for $\epsilon =0$ 
is the product of that for the equation ${\cal F}=0$ 
and that for the equation (\ref{char}).
Therefore taking the mode solution $\delta(t)=\exp(\lambda t)$ we arrive at 
the stability condition ($\Re \; \lambda<0$ for all the eigenvalues)
 in the form:
\begin{eqnarray}
(D-1)H_0 \leq 0 \nonumber \\
\frac{1}{L_{22}}\,\left[L_{12}(D-1)H_0+2(D-1)L_2-L_{11}\right]>0
\label{stabi}
\end{eqnarray}
These conditions are obtained from ${\cal F}=0$, for the whole 
equation (\ref{char}) we should 
include the additional condition $\lambda=-K<0$, 
that is $(D-1)H_0>0$ which is
not compatible with the first condition in (\ref{stabi}).
Accordingly, all de Sitter solutions are unstable, this
property of the de Sitter solutions 
doesn't depend on the dimension or on the choice of the parametrization.  

Returning to the general case with dilaton field, we will consider the 
non trivial solutions
in (\ref{16}) and (\ref{solu2}) which satisfy $K=0$. Then equations 
(\ref{stab}), (\ref{stab1}) reduce to:
\begin{eqnarray}                             
\dot \delta(-L_{11}+2(D-1)L_2)+\frac{d}{dt}\ddot \delta L_{22}+\epsilon(D-1)L_3
+\dot \epsilon(-L_{13}-2L_2)+\ddot \epsilon L_{23}=0 \nonumber \\
\delta(-2L_1 -(D-1)L_3)+\dot \delta(-2L_2-L_{13})-\ddot \delta L_{23}
-\dot\epsilon L_{33}=0
\end{eqnarray}

As before we consider the mode solutions: $\delta=\exp(\lambda t)$ and 
$\epsilon=\alpha \exp(\lambda t)$, 
then the above two equations reduce to a single fourth order 
algebraic equation. The form
of this equation when we especialize to the lagrangian expression 
in (\ref{24}) (that also
includes the case without torsion (\ref{9}))  is the following:
\begin{eqnarray}
\lambda^4+\frac{\lambda^2}{2c_4}\left(C+2(D-1)B-\frac{B^2}{2}\right)+
\frac{\lambda}{2c_4}\left(-2Bv(D-1)+\frac{AB}{4}\right)+\frac{v(D-1)A}{2c_4}=0
\label{alg}
\end{eqnarray}
where:
$$A=-4c_1 H_0 -8c_5 H_0^3+8(D-1)v$$
$$B=c_2+c_3H_0^2$$
$$C=-2c_1-12c_5 H_0^2$$
The stability in this case imposes  the signs of the coefficients 
in (\ref{alg}) to be alternate, which implies the following constraints:
\begin{eqnarray}
\frac{1}{2c_4}\left(C+2(D-1)B-\frac{B^2}{2}\right)>0\nonumber \\
\frac{1}{2c_4}\left(-2Bv(D-1)+\frac{AB}{4}\right)\leq 0 \nonumber \\
\frac{v(D-1)A}{2c_4}>0
\end{eqnarray}
In the particular case without torsion (\ref{9}), it is possible to 
get general results 
from these equations thus: if $\gamma_1>0$ and $H_0 \geq 0$ then 
the solution is stable, otherwise
it is unstable.

The study of stability of the de Sitter solutions with torsion 
(\ref{solution}) is more involved  
and it requires to consider 
the particular parametrizations and dimensions. 
However, the above three conditions are proportional to $1/c_4$, and $c_4$
only depends on $\gamma_1$, moreover the dependence on $\gamma_1$ only enters
through $c_4$ and accordingly we can conclude that for a given dimension, 
the stability 
of the solutions with dilaton and 
torsion depends on the parametrization, thus a
stable solution for certain $\gamma_1$ would be unstable if we change the 
parametrization to $-\gamma_1$. Thus for the given space-time dimension 
$D$ one can always choose the parametrization in such a way that the 
de Sitter solutions are unstable.


\baselineskip 16pt

\vskip 8mm
\noindent
{\large \bf 7.$\,$ Conclusion}

\vskip 2mm
\baselineskip 16pt

We have explored the problem of the inflationary de Sitter solutions in
higher derivative string-inspired theories of gravity with
dilaton field and with torsion. It is shown that in the absence of torsion,
the dilaton-free theory
may have de Sitter solutions only for $D\neq 4$, whereas the theory with
dilaton possesses such a solution (\ref{16}) with real Hubble parameter for
negative $\ga_3$. The lack of (anti-)de Sitter solutions for $D=4$ in the
dilatonless theory can be explained by the fact that the crucial
Gauss-Bonnet-like term becomes a total divergence in this dimension
and doesn't contribute to the equations of motion. When the
dilaton field is introduced, $D=4$ doesn't have special
features, and in particular there is a real inflationary de Sitter solution
with dilaton for $\ga_3 < 0$.

It is easy to see that the existence of (anti-)de Sitter
solutions essentially depends on the parametrization of the metric
in the target space. This is because the reparametrization (\ref{repar})
 can modify
the coefficients $a_2$ and $a_3$ and therefore change the sign of
$\ga_3$ which is essential both for the torsionless case (\ref{16})
and also in the expressions (\ref{solution}) and (\ref{solu2}) for the
solutions with torsion.

For the special "ghost-free" parametrization in the theory without torsion,
the sign of $\ga_3$
is completely determined by the sign of $a_1$ which is (in the Minkowski
signature) positive for both bosonic and heterotic string (where
$a_1 = 1/4,\,\,1/8$ correspondingly.)
\cite{mets,ket}.
Therefore in this special
parametrization there are no inflationary de Sitter solutions, whereas
they
can exist in other parametrizations. This gives one more illustration
to the strong parametrization dependence of the physical properties of
the higher order corrections to the string effective action \cite{ovrut}.

On the other hand, considering the torsion terms in the action,
for non-dynamical dilaton, unstable de Sitter solutions
appear for $D=4$. In the general case
with dilaton, we obtain  the explicit expression for
the (anti-)de Sitter solutions. 

The investigation of the stability of the (anti-)de Sitter solutions shows 
that generally this property depends on the particular dimensions and
also on the parametrization of the target-space metric. Only in the case
of the dilatonless and torsionless theory, one can drow definite conclusion
about the universal nonstable nature of the de Sitter solutions. For the 
torsionless theory with dilaton it proves possible to construct the conditions
of stability for the de Sitter solutions explicitly, those do not depend 
on dimension but only on the parametrization of the metric.
In the general case with torsion and dilaton the stability of the 
de Sitter solutions depends on the 
parametrization and dimension, but for the given dimension one can always 
choose parametrization in such a way that the solutions are unstable.

\vskip 5mm
{\bf Acknowledgments}

The authors are grateful to the referee for the suggestion to study the 
stability of the de Sitter solutions, presented in section 6.

A.L.M. thanks Prof. A. Dobado for useful discussions and also acknowledges 
partial
support by the Ministerio de Educaci\'on y
Ciencia (Spain) (CICYT AEN96-1634).

I.L.Sh. thanks the
Departamento de Fisica -- ICE, Universidade Federal de Juiz de
Fora for warm hospitality.
The work of I.L.Sh. has been
supported in part by Russian Foundation for Basic
Research under the project No.96-02-16017.


\newpage

\begin {thebibliography}{99}

\bibitem{ste} Stelle K.S.{\sl Gen.Rel.Grav.}{\bf 9} (1978) 353;
{\sl Phys.Rev.} {\bf 16D} (1977) 953.

\bibitem{gsw} Green M.B., Schwarz J.H. and Witten E.,
{\it Superstring Theory} (Cambridge University Press, Cambridge, 1987).

\bibitem{anom}
Reigert R.Y., {\it Phys.Lett.} {\bf B134} (1984) 56;
Fradkin E.S. and Tseytlin A.A., {\it Phys.Lett.} {\bf B134} (1984) 187;
Buchbinder I.L., Odintsov S.D., Shapiro I.L., {\sl Phys.Lett.} {\bf B162}
(1985) 92.
Antoniadis I., Mottola E.,  {\it Phys.Rev.} {\bf D45} (1992) 2013.

\bibitem{aeli}
Shapiro I.L., Jacksenaev A.G.,  {\it Phys. Lett.}{\bf B324}(1994) 284;
Elizalde E., Jacksenaev A.G., Odintsov S.D., Shapiro I.L.,
{\it Phys. Lett.}{\bf B328}(1994) 297.

\bibitem{book} Buchbinder I.L., Odintsov S.D., Shapiro I.L.
{\it Effective Action in Quantum Gravity}
- IOP Publishing, Bristol and Philadelphia (1992).

\bibitem{mets} Metsaev R.R., Tseytlin A.A.,  {\it Nucl.Phys.} {\bf B293}
(1987)92.

\bibitem{ket} For the list of references on the higher-loop
calculations in the non-linear sigma-models one can see [1] and
Ketov S.V., {\it Nonlinear Sigma-models in Quantum
Field Theory and Theory of Strings} -- Nauka, Novosibirsk -- (1992,
in Russian)

\bibitem{zwei} Zwiebach B.,  {\it Phys.Lett.} {\bf B156} (1985) 315.

\bibitem{dere}Deser S. and Redlich A.N.,
 {\it Phys.Lett. }{\bf B176} (1986) 350;
Fridling B.E., Jevicki A.,{\it Phys.Lett.} {\bf B174} (1986) 75;
Jones D.R.T., Lowrence A.M., {\it Z.Phys.} {\bf 42C} (1989) 153.

\bibitem{tse} Tseytlin A.A., {\it Phys.Lett.} {\bf B176} (1986) 92.

\bibitem{ovrut} Forger K., Ovrut B.A., Theisen S.J. and Waldram D.,
{\it Higher-Derivative Gravity in String Theory} -- hep-th/9605145.

\bibitem{highderi} Asorey M., L\'opez J.L. and Shapiro I.L.
{\it Some remarks on high derivative quantum gravity.}
-- DFTUZ 96/15, hep-th/9610006

\bibitem{don} Donoghue J.F., {\sl Phys.Rev.Lett.} {\bf 72} (1994) 2996;
{\sl Phys.Rev.}{\bf D50} (1994) 3874.

\bibitem{cosm1} Boulware D.G. and Deser S.,
{\sl Phys.Rev.Lett.} {\bf 55} (1985) 2656;
{\sl Phys. Lett.} {\bf B175} (1986) 409.

\bibitem{bebe} Bento M.C. and  Bertolami O.,
{\sl Phys. Lett.} {\bf B368} (1996) 198.

\bibitem{barrow} Whitt B, Phys.Lett. {\bf B145} (1984) 176;
Barrow J.D. and Ottewill A.C., {\it J.Phys.}
{\bf A16} (1983) 2757;
Barrow J.D. and Cotsakis S.,{\it Phys.Lett.} {\bf B214} (1988) 515;
H.-J.Schmidt, Class.Quantum Grav. {\bf 6} (1989) 557.

\bibitem{cosm2} Dobado A. and L\'opez A.
 {\it Phys. Lett.}{\bf B316}(1993) 250;
Dobado A. and Maroto A.L.,{\sl Phys.Rev.} {\bf D52} (1995) 1895

\bibitem{barrowmadsen}  Barrow J.D. and Madsen M., {\it Nucl.Phys.}
{\bf B323} (1989) 242.

\bibitem{Vene} Veneziano G., {\it Phys.Lett} {\bf B265} (1991) 287.

\end{thebibliography}
\end{document}